\documentclass{elsart} 
\input epsf

\begin{document}
\begin{center}  
\title{Retrieving Nuclear Information from Protons 
Propagating through a Thick Target}
\author{B.G. Giraud }  
\address{giraud@dsm-mail.saclay.cea.fr, Service de Physique Theorique,}  
\address{DSM, CEA Saclay, F-91191 Gif-sur-Yvette, France}
\author{Lon-chang Liu}  
\address{liu@lanl.gov, Theoretical Division, Group T16, MS B243,}  
\address{Los Alamos National Laboratory, Los Alamos NM 87545 USA}
\end{center} 

\begin{center}  
{\bf Abstract} 
\end{center}
The multiple scattering of high-energy particles in a thick target is 
formulated in an impact parameter representation. A formalism 
similar but not identical to that of Moli\`ere is obtained.
We show that calculations of particle beam broadening due to multiple 
Coulomb scattering alone can be given in closed form. 
The focus of this study is on whether or not   
the broadening of the Coulomb angular distribution 
prevents the retrieval of nuclear-interaction information from 
measuring the angular distributions of charged particles 
scattered from a thick target. For this purpose,
we study multiple scatterings with both the nuclear
and Coulomb interactions included and we do not make a small-angle  
expansion. Conditions for retrieving nuclear information
from  high-energy protons propagating through a block of material 
are obtained.      

\vspace{-0.cm}


\pagebreak
\section{Introduction and basic formalism}
 
Understanding and calculation of the broadening of a particle beam when
it propagates through a block of materials are important not only to make
multiple-scattering corrections for cross-section measurements in physics
experiments but also to many applications such as radiography by means of
high-energy protons.  Many different theories of the multiple scattering
of electrons by thick targets have been formulated in the past sixty years
\cite{Moli}-\cite{Beth}. The theory of Moli\`ere \cite{Moli}\cite{Beth} has
received extensive attention because it gives the best agreement  
with data concerning the broadening of Coulomb angular distribution, arising
from the multiple scattering of charged particles from the atoms in thick 
targets. An excellent and succinct derivation of the Moli\`ere 
theory was given by Mott and Massey \cite{Mott}.
 
As one deals with a large (almost astronomical) number of scatterers in a 
thick target, the numerical aspect of the calculation becomes extremely 
demanding. One can obtain good results provided that careful approximations  
are carried out. Over the years, successful parametrizations of the  
broadening of angular distributions due to Coulomb multiple scattering
have been established\cite{EJP}.\  For hadronic projectiles, such as 
protons, nuclear interactions also contribute to multiple scattering.
However, this latter aspect has not yet received sufficient attention  
in the literature.    
In this work, our focus is, therefore, on effects of multiple scattering on 
angular distributions in the Coulomb-nuclear interference region  and in the 
region where the nuclear interaction dominates. 
We formulate the multiple-scattering problem in such a way that  
not only in the case with Coulomb multiple scattering alone
can analytical evaluations of the beam broadening become possible, but also 
the calculations are greatly facilitated when both nuclear and Coulomb 
interactions are taken into consideration.  

For a very high energy (e.g., $\geq$ 20 GeV) proton scattering from a
single nucleus, the Coulomb cross section decreases rapidly with the increase 
of scattering angles in such a way that the Coulomb cross section is
negligible with respect to the nuclear cross section already at scattering
angles as small as several milliradians. However, the Coulomb peak is
rapidly broadened by proton-atomic nucleus multiple scatterings. Clearly,  
for the purpose of extracting the forward amplitude of the basic 
hadron-nucleus strong interaction, one should use targets as thin as possible  
and then employ Moli\`ere-type theory to correct for  
the Coulomb contribution 
from the measurement\cite{Shen}. 
On the other hand, in many practical applications the thickness of
the ``target'' is often fixed by specific needs, which
is by no means thin.  It becomes,
therefore, interesting to know how much nuclear information can still be  
learned from hadrons scattered from a thick target. Certainly, the 
feasibility of   
learning nuclear information can bring added value to probing materials   
with hadronic beams. 
In other words, will the broadening of
the Coulomb angular distribution render impossible any study of the nuclear
cross sections? One naturally expects that the survival of nuclear
information, if any, depends on the target thickness, i.e., on the number
of nuclear scatterers that a proton encounters in a block of material. 
We use our formalism to examine this question. 

In this section, after deriving the basic multiple-scattering formalism,   
we discuss the important bearing of an ability to retrieve 
nuclear information on many applications. In Section II, we show the 
broadening of angular distributions by Coulomb multiple scattering in closed
form. The broadening of angular distributions by combined Coulomb and nuclear 
interactions are studied by means of  semi-analytical models in regions of  
small momentum transfers (Section III) as well as large momentum transfers 
(Section IV). We find that it is possible to retrieve      
nuclear information from protons scattering from a thick 
target. Conclusions and suggestions are presented in Section V.   
    
It is well known that high-energy elastic scattering is basically forward
peaked, which allows to a very good approximation to neglect the
longitudinal momentum transfer. Glauber \cite{Glau} has shown that, in an
impact-parameter representation, every function $O(q)$ of the tranverse 
momentum transfer $q,$ whether $O$ is an amplitude or a cross-section, can be 
parametrized in terms of a profile
function $\phi(b)$ defined in the impact plane, 
\begin{equation}
O(q)=  (2 \pi)^{-2}
\int d \vec b \, \exp(i \vec q \cdot \vec b )\, \phi(b)
= (2\pi)^{-1} \int_0^{\infty} b\, db\, J_0(qb)\, \phi(b),
\label{btoq}
\end{equation}
where  azimuthal symmetry is assumed, $J_0$ is the Bessel function of the
first kind, and $q$ and $b$ are the moduli of the transverse momentum
transfer $\vec q $ and impact parameter $\vec b,$ respectively.
Nonessential complications ({\em e.g.} spins, etc) are here understood.
Conversely, the profile function results from the inverse Fourier transform
and is given by
\begin{equation}
\phi(b)= \int d \vec q \, \exp(-i \vec q \cdot \vec b )\, O(q)
= 2\pi \int_0^{\infty} q\, dq\, J_0(qb)\, O(q).
\label{qtob}
\end{equation}
 
Without loss of generality, we consider a target which consists of one kind
of nuclei. For a thin target of thickness $t$ and atomic density $\rho$ the
probability that a beam particle undergoes a scattering is
$p_1^{tot}=t\, \rho\, \sigma_1^{tot},$ where $\sigma_1^{tot}$ is the total 
cross section \cite{Joac}. (The subscript $1$ denotes the single scattering.)
The transmission probability is, therefore, given by
$p_1^{trans}=1-p_1^{tot}.$ From the definition of the differential 
cross section $\sigma_1(\vec q),$ one obtains the sum rule
$\sigma_1^{tot}=\int d\vec q\ \sigma_1(\vec q).$ 
The scattering probability density $p_1(\vec q)$ is related to the 
differential cross section by~\cite{Joac},\cite{Rodb} 
$p_1(\vec q)=t\, \rho\, \sigma_1(q)$. 
It is the probability that a particle experiences scattering in the direction
$\vec q.$ The sum rule of $\sigma_1(\vec q)$ leads to the sum rule   
$p_1^{tot} =\int d\vec q\ p_1(\vec q)$.  
Most often, only the 
modulus $q$ counts, $\sigma_1(\vec q) = \sigma_1(q)$. Hence,  
$\sigma_1^{tot}= \pi \int d(q^2)\, \sigma_1(q).$ 
 
In thick targets the beam can bounce forward from many nuclei
and/or electronic clouds and/or different atoms. These multiple
scatterings are incoherent because the scatterers are separated far apart
with respect to the ranges of the screened Coulomb and nuclear interactions
so that the scattering waves are already in the asymptotic region before
the next collision occurs. Furthermore, the target is not crystalline on a
macroscopic scale, and thus the distance between scatterers are largely
random. One must, therefore, add probabilities (not amplitudes) coming from
individual scatterings. If one splits the thick target with thickness $T$
into a large number $N$ of thin targets each with thickness $T/N,$ then
the total multistep probability for a particle to be transmitted without
any scattering is 
$P_M^{trans} \equiv P_0=\left( 1 - \sigma_1^{tot} T \rho / N \right)^N.$
 
The differential probability density for just one scattering in this
 situation with many thin targets, each with thickness $T/N,$ is
\begin{equation}
P_1(\vec q)=N\ \sigma_1(\vec q)\ (T/N)\ \rho\
\left(1-\sigma_1^{tot}T\rho/N\right)^{N-1}.
\end{equation}
This represents a combination of scattering from any one single layer and
transmission through the remaining $(N-1)$ layers without scattering.
The factor $N$ in front of the right hand side accounts for the $N$ layers,
obviously.
 
For double scattering one must count pairs of layers and fold
two single-scattering probability densities $p_1$. Hence, 
\begin{equation}
P_2(\vec q)=\frac{N(N-1)}{2}\ \left(\frac{T\rho}{N}\right)^2\ 
\sigma_2(\vec q)\ \left[1-\frac{\sigma_1^{tot}T\rho}{N}\right]^{N-2},
\end{equation}
where
\begin{equation}
\sigma_2(\vec q)=\int d \vec q^{\ \prime}\
\sigma_1(\vec q^{\ \prime})\
\sigma_1(\vec q -\vec q^{\ \prime}).
\end{equation}
 
For triple scattering, an identical argument induces the result
\begin{equation}
P_3(\vec q)=\left(\matrix{N \cr 3} \right)\, \left(\frac{T\rho}{N}\right)^3
\, \sigma_3(\vec q)\, \left[1-\frac{\sigma_1^{tot}T\rho}{N}\right]^{N-3},
\end{equation}
where
\begin{equation}
\sigma_3(\vec q)= \int d \vec q^{\ \prime} \int d \vec q^{\ \prime \prime}\
\sigma_1(\vec q^{\ \prime})\
\sigma_1(\vec q^{\ \prime \prime}-\vec q^{\ \prime}) \
\sigma_1(\vec q-\vec q^{\ \prime \prime})
\end{equation}
is a double convolution. Again, the first factor is the counting of all
triplets of layers.
 
It is useful at this stage to introduce the profiles
\begin{equation}
\phi_1(b)= \int d \vec q \, \exp (-i \vec q \cdot \vec b )\,
\sigma_1(\vec q) = 2\, \pi\, \int_0^{\infty}q\, dq\, J_0(qb)\, \sigma_1(q),
\label{frstep}
\end{equation}
with $\phi_1(0)=\sigma_1^{tot},$ and, $\forall n,$
\begin{equation}
\phi_n(b)= \int d \vec q \, \exp (-i \vec q \cdot \vec b )\ 
\sigma_n(\vec q) = \left[\phi_1(b)\right]^n.
\label{convolstep}
\end{equation}
If we define $\Phi_1$ as the profile of $P_1,$ then we see that,
\begin{equation}
\Phi_1(b)=N \left(\frac{T \rho}{N}\right) \phi_1(b) \
\left(1-\sigma_1^{tot}T\rho/N\right)^{N-1},
\end{equation}
and, more generally, the profile of $P_n$ reads,
\begin{equation}
\Phi_n=
\left(\matrix{N \cr n} \right)\, \left(\frac{T\, \rho\, \phi_1}{N}\right)^n
\, \left[1-\frac{\sigma_1^{tot}\, T\, \rho}{N}\right]^{N-n} .
\end{equation}
 
The ``total multistep profile'' due to $P_M=\sum_{n=1}^{N} P_n$ can be
written as,
\begin{eqnarray}
\Phi_M(b) & =& \left(1 - \frac{\sigma_1^{tot}\, T\, \rho}{N} +
\frac{T\, \rho\ \phi_1(b)}{N}\right)^{N} - P_0 \nonumber \\
 & =& 
\left\{1 + \frac{\nu\ [ \phi_1(b)/\sigma_1^{tot}-1 ]}{N}\right\}^{N} - 
\left(1-\nu/N\right)^N, 
\label{prelimit}
\end{eqnarray}
where we have introduced the parameter $\nu=\sigma_1^{tot} T \rho.$ 
\ Like $p_1^{tot},$ \ $\nu$ is dimensionless. However, it is not a 
probability for a thick target. Indeed, because the mean free  
path, $\Lambda$, of a  
beam particle is $1/(\rho\sigma_1^{tot})$ ,  
therefore, $\nu = T/\Lambda$ represents the average number of 
collisions of a beam particle when it passes through a 
target of thickness $T$. Hence $\nu$ can be a very large number.   

In the limit, $N\rightarrow \infty$, we obtain,
\begin{equation}
\Phi_M(b)=
\exp\left[ \nu\, \left(\frac{\phi_1(b)}{\sigma_1^{tot}}-1\right) \right]
- \exp(-\nu)\ .  
\label{thrdstep}
\end{equation}
Here, we emphasize the {\it nonlinear} action of $\nu$ on $\Phi_M(b)$.   
The second term in Eq.(\ref{thrdstep}) comes from the
limit of $P_0$, it also shows that 
$\nu$ has the meaning of a beam decay rate in a target of thickness $T$.   
Note that this latter term is neglected in the
Moli$\grave{e}$re theory \cite{Moli} as well as in its reformulation by
Bethe \cite{Beth} and by Mott and Massey \cite{Mott}.
Hence, we find a form similar but not identical to that of the
Moli$\grave{e}$re theory for the final, multistep probability density
$P_M:$
\begin{equation}
P_M(q)=\ (2 \pi)^{-2}\, \exp(-\nu)\, \int d \vec b \,
\exp (i \vec q \cdot \vec b )\,
\left\{\exp\left[ \nu\, \frac{\phi_1(b)}{\sigma_1^{tot}} \right]-1\right\}
\,.
\label{backagain}
\end{equation}
Again, we note that $P_M(q)$ depends {\it nonlinearly} on $\nu$. 
It is appropriate to underline the importance
of the above-mentioned extra term in the present formulation.
If $T$ is small rather than large,  $T=t,$ then $\nu$ is
also small. The presence of this extra term reduces $P_M$ to $p_1=t\, \rho\,
\sigma_1,$ as should be for thin targets. On the other hand,  
$\phi_1(b)\rightarrow  0$ as $b\rightarrow \infty$. 
Consequently, the first term in the curly bracket has a limit, 
$\exp[\ ] \rightarrow 1,$ that exactly compensates the second term in the 
curly bracket, ensuring the convergence of the integration for $P_M(q)$.
 
It is interesting to note from Eq. (\ref{backagain}) that    
\begin{equation}
P_M^{tot} \equiv \int d \vec q\ P_M(\vec q)=1-\exp(-\nu)=1-P_{Mt}\ ,  
\end{equation}
where $P_{Mt}$ is the total transmission probability. This last equation is a
sum rule for the multiple-scattering probability. 
Multiple-scattering differential cross sections
can be related to probability densities by the general relation 
\begin{equation} 
P_M(q)=T\, \rho\,\sigma_M(q)\ , 
\label{general1}
\end{equation} 
in the same way $\sigma_1(\vec q)$ is to $p_1(\vec q)$.  
In summary, three steps thus occur in this formalism: {\it (i)} Fourier
transform thin target data $\sigma_1$ into their profile $\phi_1,$ see
Eq. (\ref{frstep}); {\it (ii)} Find $\nu=T\, \rho\, \sigma^{tot}_1$ and
exponentiate $T\, \rho\, \phi_1,$ see Eq. (\ref{thrdstep}); 
{\it (iii)} Fourier transform $\Phi_M$ back into a probability distribution 
$P_M,$ see Eq. (\ref{backagain}).

It is reasonable to parametrize the single-scattering 
distribution as $\sigma_1 = \sigma_{1c}+\sigma_{1n}$ with $\sigma_{1c}$ being 
the Coulomb cross sections and $\sigma_{1n}$ the sum of cross sections of 
nuclear scattering and   
nuclear-Coulomb interference. 
The separation of $\sigma_1$ into $\sigma_{1c}$ and $\sigma_{1n}$ induces
the same for the profile function: $\phi_1 = \phi_{1c} + \phi_{1n}.$ The
relation
\begin{equation}
\exp(\beta\phi_{1c})\, \exp(\beta\phi_{1n})-1=[\exp(\beta\phi_{1c})-1]
+\exp(\beta\phi_{1c})\,
[\exp(\beta\phi_{1n})-1]
\end{equation}
with $\beta = \nu/\sigma_1^{tot}$ 
then leads to a split of $P_M$ as the sum of two probability densities,
\begin{equation}
P_{Mc}=\ \frac{\exp(-\nu)}{(2\pi)^2} \int d \vec b \,
\exp (i \vec q \cdot \vec b ) \left\{ \exp\left[ \nu\,
\frac{\phi_{1c}(b)}{\sigma_1^{tot}} \right]-1
\right\}
\label{forward}
\end{equation}
and
\begin{equation}
P_{Mn}=\ \frac{\exp(-\nu)}{(2\pi)^2} \int d \vec b \,
\exp (i \vec q \cdot \vec b ) \, \exp\left[ \nu
\frac{\phi_{1c}(b)}{\sigma_1^{tot}} \right]\,
\left\{\exp\left[ \nu\, \frac{\phi_{1n}(b)}{\sigma_1^{tot}} \right]-1
\right\} .
\label{correctionN}
\end{equation}
This allows a perturbative consideration of nuclear effects at those
angles where Coulomb scattering dominates. Note that the exponent in
Eq. (\ref{forward}) contains a denominator $\sigma_1^{tot}$ and not
$\sigma_{1c}^{tot}.$ Hence $P_{Mc}$ is proportional to a pure Coulomb process
with an effective value of $\nu,$ namely
$\nu_{eff}=(\sigma_{1c}^{tot}/\sigma_1^{tot})\, \nu.$

Alternately, at angles where nuclear scattering might dominate, the roles
of $\phi_{1c}$ and $\phi_{1n}$ can be interchanged to generate similar
formulae, namely,  
\begin{equation}
P^{\, \prime}_{Mn}=\ \frac{\exp(-\nu)}{(2\pi)^2} \int d \vec b \,
\exp (i \vec q \cdot \vec b ) \, \left\{\exp\left[ \nu\,
\frac{\phi_{1n}(b)}{\sigma_1^{tot}} \right]-1
\right\}
\label{backward}
\end{equation}
and
\begin{equation}
P^{\, \prime}_{Mc}=\ \frac{\exp(-\nu)}{(2\pi)^2} \int d \vec b \,
\exp (i \vec q \cdot \vec b ) \, \exp\left[ \nu
\frac{\phi_{1n}(b)}{\sigma_1^{tot}} \right]\,
\left\{\exp\left[ \nu\, \frac{\phi_{1c}(b)}{\sigma_1^{tot}} \right]-1
\right\}\ .
\label{correctionC}
\end{equation}
One can parametrize the screened Coulomb interactions  as
$\sigma_{1c}(q)=\sum_{\alpha} C_{\alpha}\, ( q^2 + 
\kappa_{\alpha}^2 )^{-\alpha}.$ 
The powers $\alpha,$ screening momenta $\kappa_{\alpha}$, and normalizations
$C_{\alpha}$ are mainly functions of the charge $Z$ of each individual
nucleus. But $\sigma_{1n}$ will depend on both $Z$ and the mass number $A.$
The global normalization of $\sigma_M$ will also depend on the target
thickness or the parameter $\nu$.
Hence, the theory is essentially driven by three parameters of a thick
target, namely, $Z,$ $A$ and $\nu.$ Experimental measurements of $\sigma_M$
might, conversely, permit a determination of such three parameters when the
nuclear nature of the target is unknown {\it a priori}, as is most often
the case for radiographic studies where a recovery of $A,$ besides $Z,$
would be precious. Success will occur, however, only if multiple scattering
does not spoil the information carried by $A.$ This question is the main
concern of the following sections.

\section{Broadening of angular distributions by multiple scatterings} 
 
The mean-square width $\langle q^2 \rangle_1$ of the distribution $\sigma_1$
represents a useful observable for the broadening of the cross-section
distribution $\sigma_1$ and can be defined by an integral
\begin{equation}
\langle q^2 \rangle_1=
(\sigma_1^{tot})^{-1}\, \int d\vec q\ q^2\, \sigma_1(\vec q)=
2\, \pi\, (\sigma_1^{tot})^{-1}\, \int_0^{\infty} dq\ q^3\, \sigma_1(q), 
\end{equation}
if it converges. The use of Eq.(\ref{frstep}) and elementary properties of
the Fourier transform allows us to write  
\begin{equation}
\langle q^2 \rangle_1 = - \lim_{b\, \rightarrow\, 0}
\, \left(\frac{1}{b}\, \frac{d}{db}\, b\, \frac{d}{db}\right)\, 
\frac{\phi_1(b)}{\phi_1(0)}\, ,
\end{equation}
where the operator between brackets $(\ )$ comes from a two-dimensional 
Laplacian in cylindrical coordinate space. The same procedure gives
the mean-square width of $\sigma_M(\vec q)$ as  
\begin{equation}
\langle q^2 \rangle_M= - \, \lim_{b\, \rightarrow\, 0}
\, \left(\frac{1}{b}\, \frac{d}{db}\, b\, \frac{d}{db}\right)\, 
\frac{\Phi_M(b)}{\Phi_M(0)}\, , 
\label{ruse}
\end{equation}
where $\Phi_M(0)= (1 - e^{-\nu})$. Assume, for the sake of the argument,
that $\sigma_1(q)$ is a Gaussian, 
\begin{equation}
\sigma_1(q)=C \exp[-q^2/(2 \kappa^2)],
\end{equation}
where $C$ is a suitable normalization and the parameter $1/\kappa$ is the   
interaction range. For instance, if one discusses screened-Coulomb 
interactions, then $1/\kappa$ is of the scale of an atomic radius. Then one 
obtains
$\sigma_1^{tot}=2\, \pi\, C\, \kappa^2$\ ,  
$\langle q^2 \rangle_1=2 \kappa^2$, and 
\begin{equation}
\phi_1(b)=2\, \pi\, C\, \kappa^2\, \exp(-b^2 \kappa^2/2).
\end{equation}
From Eqs. (\ref{thrdstep}) and (\ref{ruse}) one further obtains 
\begin{equation}
\Phi_M(b)=\exp\left[\nu\left(e^{-b^2 \kappa^2/2}-1\right)\right]-\exp[-\nu]
\end{equation}
and 
\begin{equation}
\langle q^2 \rangle_M=\frac{\nu\ \langle q^2 \rangle_1}{1-\exp(-\nu)}\, .
\label{broadenrule} 
\end{equation}
Since $\nu$ is large in general, the denominator is $\simeq 1$. Consequently,
the multiple scattering has broadened the mean-square width by a factor   
$\nu$, as might be expected from a Brownian motion in the transverse-momentum
space. The multiplication of $\langle q^2 \rangle_1$ by $\nu$ also 
occurs if we start from a ``polynomial Gaussian distribution'' 
$(q/\kappa)^{2n} \exp[-q^2/(2 \kappa^2)].$  This growth rate is very 
general and can be viewed as one more version of the central limit theorem. 
As additional evidence, one finds that if $\sigma_1(q)$ is of the 
functional form $\sigma_1(q)=C (q^2+\kappa^2)^{-n}$ with an exponent 
$n>2$, then $\langle q^2 \rangle_1=\kappa^2/(n-2)$ and again
$\langle q^2 \rangle_M=\nu\, \langle q^2 \rangle_1/(1-e^{-\nu})\simeq
\nu \langle q^2\rangle_1.$
 
In the following, we illustrate the broadening of the 
cross-section distribution in the case of a screened Coulomb scattering.  
We fit $\sigma_1$ at small angles by a few terms of the sum 
\begin{equation}
\sigma_{1c}(q)=\sum_{m>2} \frac{ C_{m}}{(q^2+\kappa_{m}^2)^{m}}\, ,
\label{ansatzsum1}
\end{equation}  
where $m$ can be half-integers as well as integers, and $C_m$ and
$\kappa_m$ are fitting parameters.  
It follows that   
\begin{equation} 
\sigma^{tot}_{1c} = \pi\, \sum_m \frac{C_{m} }{ (m-1)\, \kappa^{2(m-1)} }\, .
\label{ansatzsum2}
\end{equation}
For definiteness, we take two terms with $m=5/2$ and $4$, namely,
\begin{equation}
\sigma_{1c}(q)=\frac{C_{5/2}}{(q^2+\kappa_{5/2}^2)^{5/2}}
+\frac{C_{4}}{(q^2+\kappa_{2}^2)^4}\ . 
\label{ansatzsig1c}
\end{equation}
Hence,
\begin{equation} 
\sigma^{tot}_{1c} = \frac{2\pi C_{5/2}}{3\kappa_{5/2}^{3} }  
 + \frac{\pi C_{4}}{3\kappa_4^{6} }\ . 
\label{ansatz2terms} 
\end{equation} 
Dividing both sides by $\sigma_{1c}^{tot}$, we obtain
\begin{equation} 
1 = \frac{2\pi C_{5/2}}{3\kappa_{5/2}^{3}\, \sigma_{1c}^{tot} }  
 + \frac{\pi C_{4}}{3\kappa_4^{6}\, \sigma_{1c}^{tot} } \equiv a_{5/2} +
a_4\ .
\label{ansatzsum4}
\end{equation} 
Eq. (\ref{ansatzsum4}) shows that both 
$a_{5/2}$ and $a_4$ are dimensionless numbers between $0$ and $1.$ An
advantage of using Eq. (\ref{ansatzsum1}) is that its Fourier transform  
gives the profile function $\phi_{1c}$ in terms of analytical functions which
can be easily analyzed, i.e.,    
\begin{equation}
\phi_{1c}(b)=\frac
{2 \pi C_{5/2} (1+\kappa_{5/2}b) \exp(-\kappa_{5/2}b)}{3\, \kappa_{5/2}^3} +
\frac{\pi C_4 b^3 K_3(\kappa_4 b)}{24\, \kappa_4^3},
\label{ansatzphi1c}
\end{equation}
where $K_3$ is the modified Bessel function of the second kind. One verifies
easily that  
$\phi_{1c}(0)=
2\pi C_{5/2}/(3\kappa_{5/2}^3)+\pi C_4/(3 \kappa_4^6)= \sigma^{tot}_{1c}. $ 
The ``Coulomb'' multistep profile then reads
\begin{equation}
\Phi_{Mc}=\exp\left\{\nu \left[ 
a_{5/2} (1+\kappa_{5/2} b) e^{-\kappa_{5/2} b}
 +
\frac{a_4b^3}{8\kappa_4^3} K_3(\kappa_4 b) -1 \right]\right\}-e^{-\nu}.
\end{equation}
When $b \rightarrow \infty,$ it is easy to verify that
Eq. (\ref{ansatzphi1c}) induces exponential decreases with ranges
$\kappa_{5/2}^{-1}$ and $\kappa_4^{-1}.$ Numerical integrals with such
integrands converge well. The final integral for the Coulomb cross section
then reads
\begin{equation}
\sigma_{Mc}(q)=\frac{1}{2\, \pi\, T\, \rho}\,
\int_0^{\infty} db\, b\, J_0(q b)\, \Phi_{Mc}(b).
\end{equation}

Let $\kappa$ be an average between the two momenta $\kappa_{5/2}$ and
$\kappa_4,$ which are both atomic scales. It is now convenient to scale
momenta and lengths as $q=\kappa Q$ and $b=B/\kappa.$ The dimensionless
$Q$ will be a few units or a few tens, if one wants to describe scattering 
angles moderately larger than the Coulomb peak. We also expect that the 
values of $B$ contributing to the integral  
\begin{eqnarray}
\sigma_{Mc}(Q)&=&\frac{\exp(-\nu)}{2 \pi \kappa^2 T \rho}  
 \int_0^{\infty} dB\, B\, J_0(Q B) \nonumber \\  
&\times& \left\{ \exp \left[ 
 \nu a_{5/2} \left(1+\frac{\kappa_{5/2}}{\kappa}B\right)
e^{-\kappa_{5/2}B/\kappa}   +
\frac{  \nu C_4 B^3}{8\kappa^3} K_3\left(\frac{\kappa_4 }{\kappa} B\right) 
 \right] - 1 \right\} 
\label{numericsMc} 
\end{eqnarray}
should be mainly between 0 and several units. In atomic units, all parameters
$\kappa,$ $\kappa_{5/2}/\kappa$ and $\kappa_4/\kappa$ are of order $1.$ It
remains to estimate the dimensionless magnitudes of $\nu\, a_{5/2}$ and
$\nu\, a_4/8.$ From Eq.(\ref{ansatzsum4}), $a_{5/2}$ and $a_4$ are moderate
fractions of $1.$ It is thus the large number $\nu$ that drives the integrand.

It is also convenient to write
\begin{equation}
 \sigma_1(q)  =  \kappa^{-4}\, \sigma_1(Q)\ , \ \ \ \  
 \phi_1(b) = \kappa^{-2}\, \phi_1(B)\  
\label{def2}
\end{equation}
with $\sigma_1(Q)$ and  $\phi_1(B)$ being dimensionless.  
Because we work with systems of atomic scale,    
we further set $\kappa$ to be $1,$ meaning that our primary scale is
``atomic''. In this scale, all lengths and momenta will, respectively, be 
given in units of atomic radius and its inverse. 
 
To show the shrinking of profiles by multiple scatterings we plot $\phi_{1c}$
and $\Phi_{Mc}$ in Fig. 1 as the ``crosses'' and solid curves when
$\sigma_{1c}= (Q^2+1)^{-5/2}$, and, respectively, as the ``circles'' and 
dashed curves when $\sigma_{1c}=(Q^2+1)^{-4}$. As one can see, the 
dashed and solid curves do decay faster than their respective single 
scattering partners.
\begin{figure}[htb] \centering
\mbox{  \epsfysize=120mm
         \epsffile{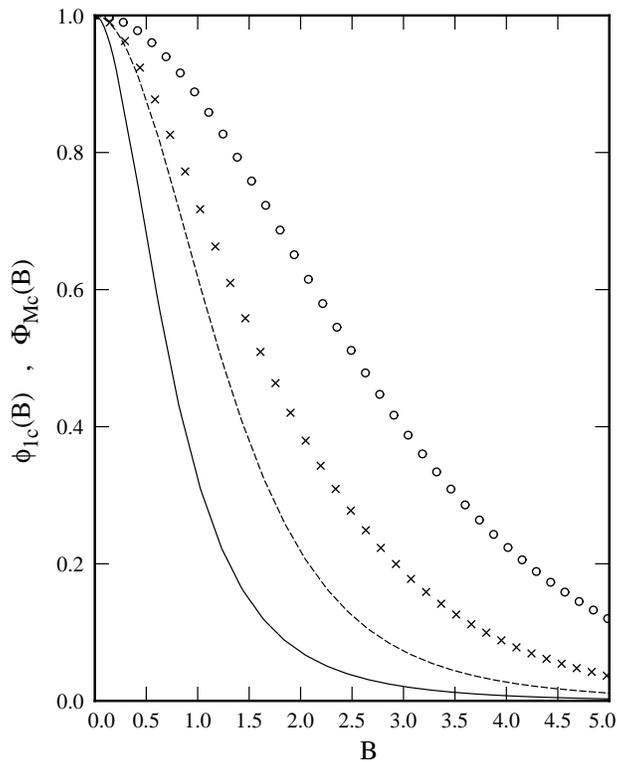}
     }
\caption{Crosses: $\phi_{1c}(B)$ for $\sigma_{1c}(Q)= (Q^2+1)^{-5/2}.$
Solid curve: the corresponding multistep $\Phi_{Mc}(B)$ if $\nu=4.$ 
Circles and dashed curve: $\phi_{1c}(B)$ and $\Phi_{Mc}(B)$
(with $\nu=4)$ for $\sigma_{1c}=(Q^2+1)^{-4}.$ All profiles normalized to 1
at $B=0$. }
\end{figure}
\begin{figure}[htb] \centering
\mbox{  \epsfysize=120mm
         \epsffile{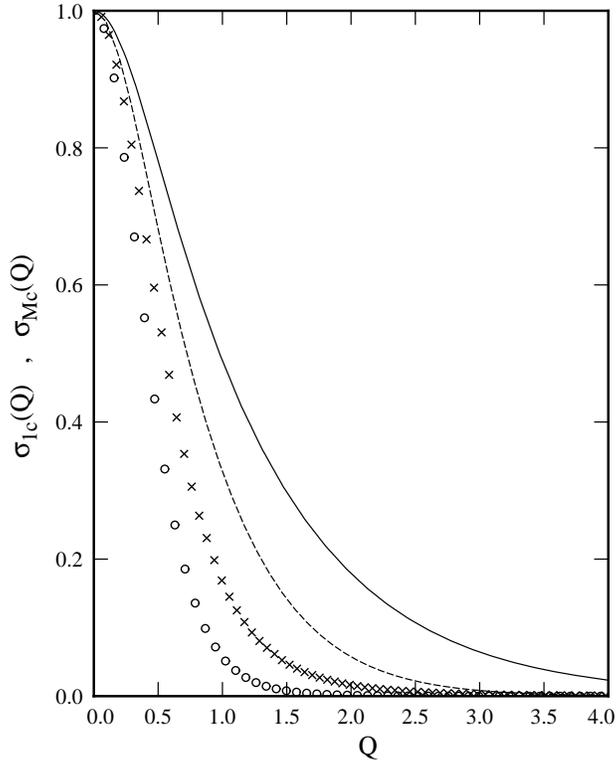}
     }
\caption{Crosses: $\sigma_{1c}(Q)=(Q^2+1)^{-5/2}.$ Solid curve: the
corresponding $\sigma_{Mc}(Q)$ for $\nu=4.$ Circles:
$\sigma_{1c}(Q)=(Q^2+1)^{-4}.$ Dashed curve: the corresponding
$\sigma_{Mc}(Q)$ for $\nu=4.$ All cross sections normalized to $1$ at $Q=0.$}
\end{figure}
\vspace{-0cm}
The Fourier transform of $\Phi_{Mc}$ then leads to the expected 
broadening of $\sigma_{Mc}$, as shown in Fig. 2. (For graphical convenience
we used $\nu=4$ in Figs. 1 and 2, which is much smaller than physical $\nu$ 
but is demonstrative enough.)

\section{Loss of nuclear information at small momentum transfers} 

In this section, we introduce a semi-realistic model for $\sigma_1(Q)$ which
contains ``nuclear'' information. We will investigate (a) changes of
normalizations and not just shrinking or dilation of shapes of $\sigma(Q)$,
and (b) how nuclear information may become lost.
We will, therefore, illustrate the blurring of signal through the study of
various relevant quantities, such as $P_M, P_{Mc}, P_{Mn}, P'_{Mc}, P'_{Mn}$.   
We also use analytical models to ensure that   
the blurring comes from 
physics and is not a result of numerical imprecision. 
A good analytical model must satisfy the following constraints:    
(i) positivity of the sum of $\sigma_{1c}$ and $\sigma_{1n}$;  (ii) big 
contrasts between maxima and minima; (iii) analyticity in both the momentum 
and the impact parameter representations; and (iv) significant differences between 
the atomic and the nuclear scales for profiles. 

Let
\begin{equation}
\sigma_{1c}(Q) = ( 1 + Q^2 )^{-2}\, ( 1 + Q^2/100 )^{-1}.
\label{mdlsig1c}
\end{equation}
This is qualitatively realistic, because the factor, $( 1 + Q^2 )^{-2},$
represents a screened Coulomb scattering. The additional, artificial factor,
$(1+Q^2/100)^{-1},$ is here just for the convergence of 
$\langle Q^2 \rangle_{1c}.$ Then we further use,
with $\sigma_{1n}(0)=0,$  the following semi-realistic $\sigma_{1n}$:
\begin{eqnarray}
\sigma_{1n}(Q) =
11\ Q^2 / 65060976287632087746874800000000\ \nonumber \\
\times  \exp(1/2 - Q^2/50)\ 
\Big[\, 24684880296681586800\ e^{\frac{12}{25}}\
       ( 3600 - 169\, Q^2 + Q^4 )^2 \nonumber \\  
- 175731507577476\, (Q^2-144)^2\, (491485925 - 20593402\, Q^2  
+\, 215573\, Q^4) \nonumber \\ 
+ 40919125\ e^{\frac{119}{50}}\ (Q^2-25)^2\,
       (3973881778272 -  
103796385841\, Q^2 + 463199137\, Q^4)\, \Big]\, 
\label{mdlsig1n}
\end{eqnarray} 
The quality of the model with respect to the requirements (ii) and (iv) are 
evidenced by Fig. 3. 

As it is allowed by the split of $\sigma_1$ into a
``Coulomb'' part ($\sigma_{1c}$) and a ``nuclear'' part ($\sigma_{1n}$), 
our $\sigma_{1n}$ can be positive or negative, as long as $\sigma_1$ remains
positive. Our $\sigma_{1n}$ was fine tuned to create four clear ``nuclear''
signals, namely two maxima of $\sigma_1$ near $Q=8$ and $18,$ and, as
signatures of interferences, two sharp minima at $Q=5$ and $12.$ Furthermore,
we adjusted its parameters so that the maxima do not exceed $\sim 1\%$ of the
forward peak of $\sigma_{1c}.$ Note also that our model $\sigma_{1n}$ has
only two maxima and, thus, carries no ``nuclear'' information for $Q > 40$.
This is designed to track whether or not the maxima, if they survive the
blurring of angular distribution by multiple scatterings, would migrate
towards larger values of $Q.$ The $\log_{10}\, \sigma_1$ and
$\log_{10}\, \sigma_{1c}$ of our toy model are shown as functions $\sigma(Q)$ 
in Fig. 3. It is trivial to deduce $\sigma_{1n}$ visually.
\begin{figure}[htb] \centering 
\mbox{  \epsfysize=120mm 
         \epsffile{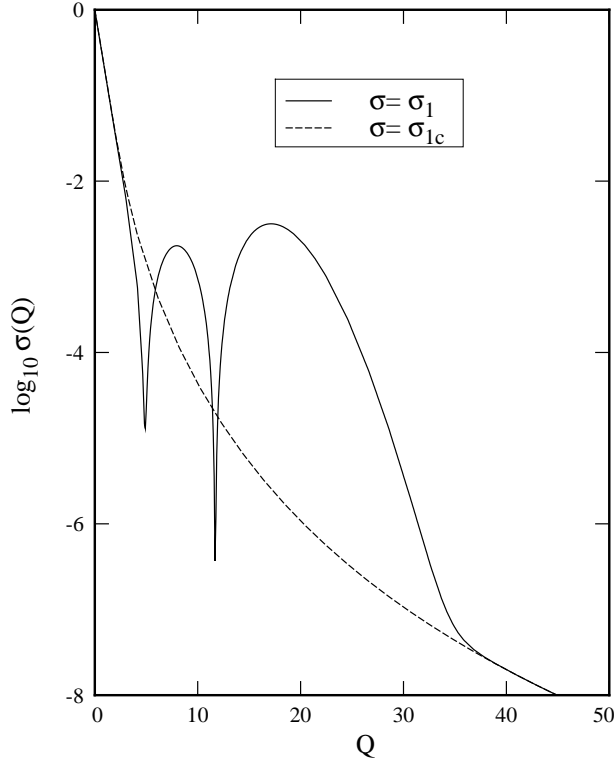} 
     }  
\caption{$\log_{10}\, \sigma_{1}$ (solid curve) 
 and $\log_{10}\, \sigma_{1c}$ 
(dashed curve) as functions of $Q$. }  
\end{figure}
\vspace{-0cm}

The corresponding profiles read, in closed forms,
\begin{equation}
\phi_{1c}(B) =
200\, \pi \, \left( \frac{-K_0(B)}{9801} + \frac{K_0(10\,B)}{9801} + 
    \frac{B\, K_1(B)}{198} \right),
\label{mdlprf1c}
\end{equation}
and, with $u=-e^{12/25}$ and $v=-e^{119/50},$
\begin{eqnarray}
 &\phi_{1n}(B) = 
 11/2081951241204226807899993600\ \pi\, \exp(1/2 - 25\, B^2/2)\, \times
\nonumber \\
&\Big[ 152587890625\, B^{10}\, (37882968282999233748 + 24684880296681586800\,
u \nonumber \\
 &+ 18953703386795125\, v) - 
4638671875\, B^8\, (1727605574138750181696 + 
\nonumber \\
 &1184874254240716166400\, u + 973536146177791625\, v) + 2734375\, B^6\, 
\times \nonumber \\
 &(1318896762458059001549772 + 935772074343960364976400\, u + 
\nonumber \\
 &815387774668679551625\, v) - 625\, B^4\, (987420216355162578690908328 +
\nonumber \\
 &716741544586342715769420000\, u + 649720700222675824315625\, v) + 
\nonumber \\
 &2800\, (12734093638656544401340038 + 9458232286390484639445000\, u 
\nonumber \\
 &+ 8603026609006638128125\, v)\, B^2 - 8\, (53363090971589265717153336 +
\nonumber \\ 
 &40407914801652923512260000\, u + 34558380311245783728125\, v) \Big]\, .
\label{mdlprf1n}
\end{eqnarray}

These profile functions are shown in Fig. 4. The width of $\phi_{1n}$ is
significantly smaller than that of $\phi_{1c},$ as one should expect when
comparing a ``nuclear'' profile to an ``atomic'' one. A geometrical ratio of
widths might be $\sim 10^{-4}$ or even $\sim 10^{-5},$ but the model ratio
we choose, between $\sim 1/5$ and $\sim 1/10,$ is sufficient for a pedagogical
study and much more convenient numerically.
\begin{figure}[htb] \centering 
\mbox{  \epsfysize=120mm 
         \epsffile{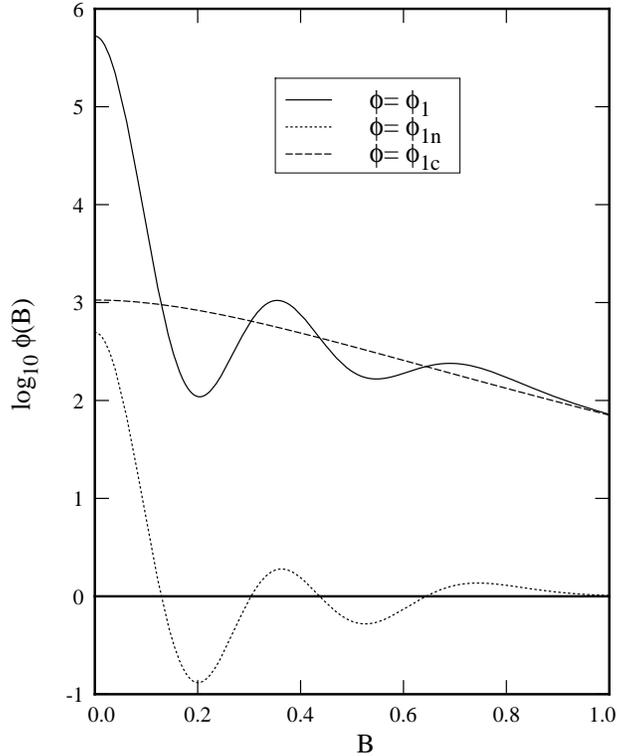} 
     }  
\caption{ $\log_{10} \phi_1$ (solid curve),  
 $\log_{10} \phi_{1n}$ (dotted curve), 
and $\log_{10} \phi_{1c}$ (dashed curve) 
as functions of $B$.} 
\end{figure}

This choice of ``data'' gives, after a numerical implementation of 
Eq. (\ref{backagain}), the total multistep probability distributions 
shown in Fig. 5. The solid curve is the same as that in Fig. 3, namely,
$\log_{10}\, \sigma_1.$ The dashed, linked-crosses, and linked-circles curves
represent $\log_{10}\, P_M$ for $\nu=4,$ $9$ and $16,$ respectively. The 
result is striking, on two counts: {\it (i)} the forward peak is more and more 
damped, the distributions extending more and more towards larger momenta, and
{\it (ii)} the ``nuclear information'', whether minima or maxima, becomes 
rapidly blurred beyond recognition. Furthermore, the broadening of
distributions does not seem to push much residual information towards 
larger momenta.
\begin{figure}[htb] \centering 
\mbox{  \epsfysize=120mm 
         \epsffile{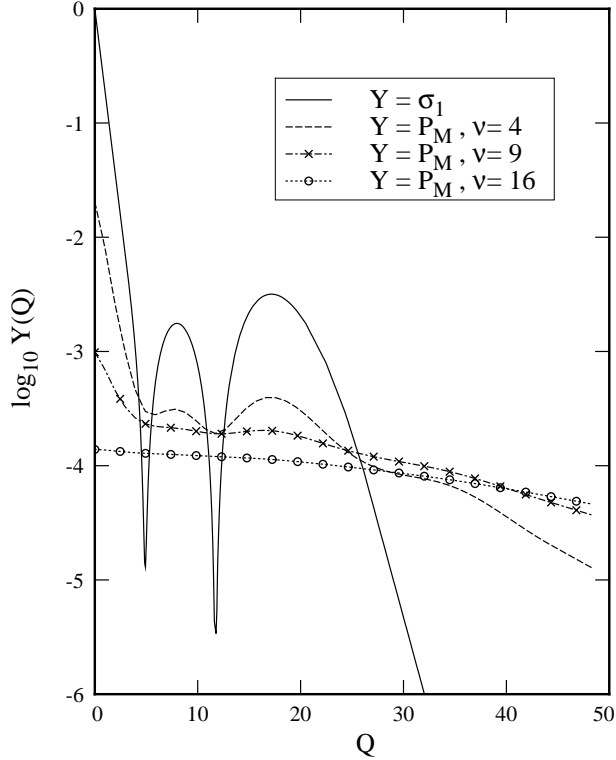} 
     }  
\caption{The dependences on $Q$ of $\log_{10}\, \sigma_1$ of single
scattering (solid curve) and $\log_{10}\, P_M$ of multiple scattering   
when $\nu=4$ (dashed curve), $9$ (linked-crosses curve), 
and $16$ (linked-circles curve), respectively.}
\end{figure}
\vspace{-0.5cm}
The broadening process is also confirmed by the behavior of the component 
$P_{Mc}$ of $P_M$, shown in Fig. 6. 
\begin{figure}[htb] \centering 
\mbox{  \epsfysize=120mm 
         \epsffile{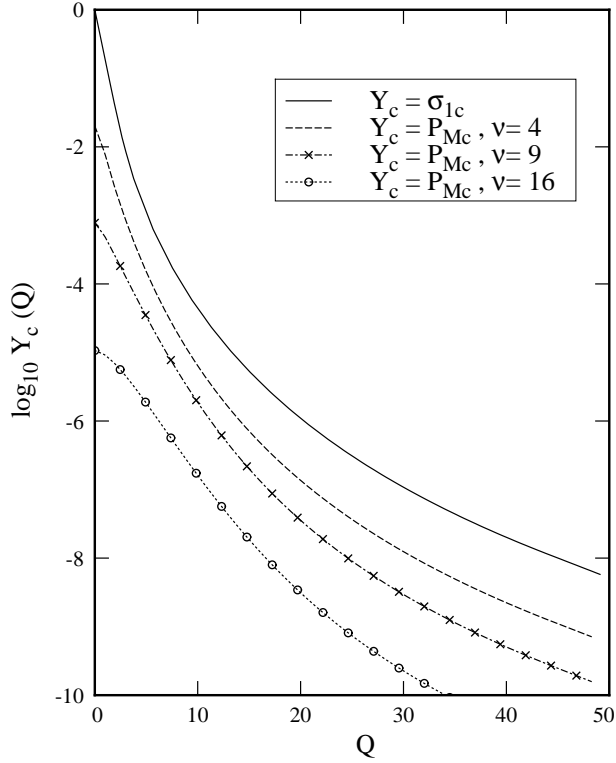} 
     }  
\caption{Solid curve: 
$\log_{10}\, \sigma_{1c}(Q).\ $ Dashed, linked-crosses and linked-circles 
curves:  $\log_{10}\, P_{Mc}(Q)$ for $\nu=4,$ $9$ and $16,$ respectively.}
\end{figure}
In our model $\sigma_1^{tot} \simeq 5.7$
and $\sigma_{1c}^{tot} \simeq 3.0.$  We chose a large nuclear contribution,
$\sigma_{1n}^{tot} \simeq 2.7,$ in order to emphasize nuclear effects.
However, at $Q<40,$ even this exaggerated nuclear information did not survive
multiple scatterings. 
 
In Fig. 7, we show the various probability distributions $P_{Mn}(Q)$. We note
again that multiple scatterings wash away nuclear information. A similar
feature is also seen in the $P^{\, \prime}_{Mn}$ given in Fig. 8.
\begin{figure}[htb] \centering 
\mbox{  \epsfysize=120mm 
         \epsffile{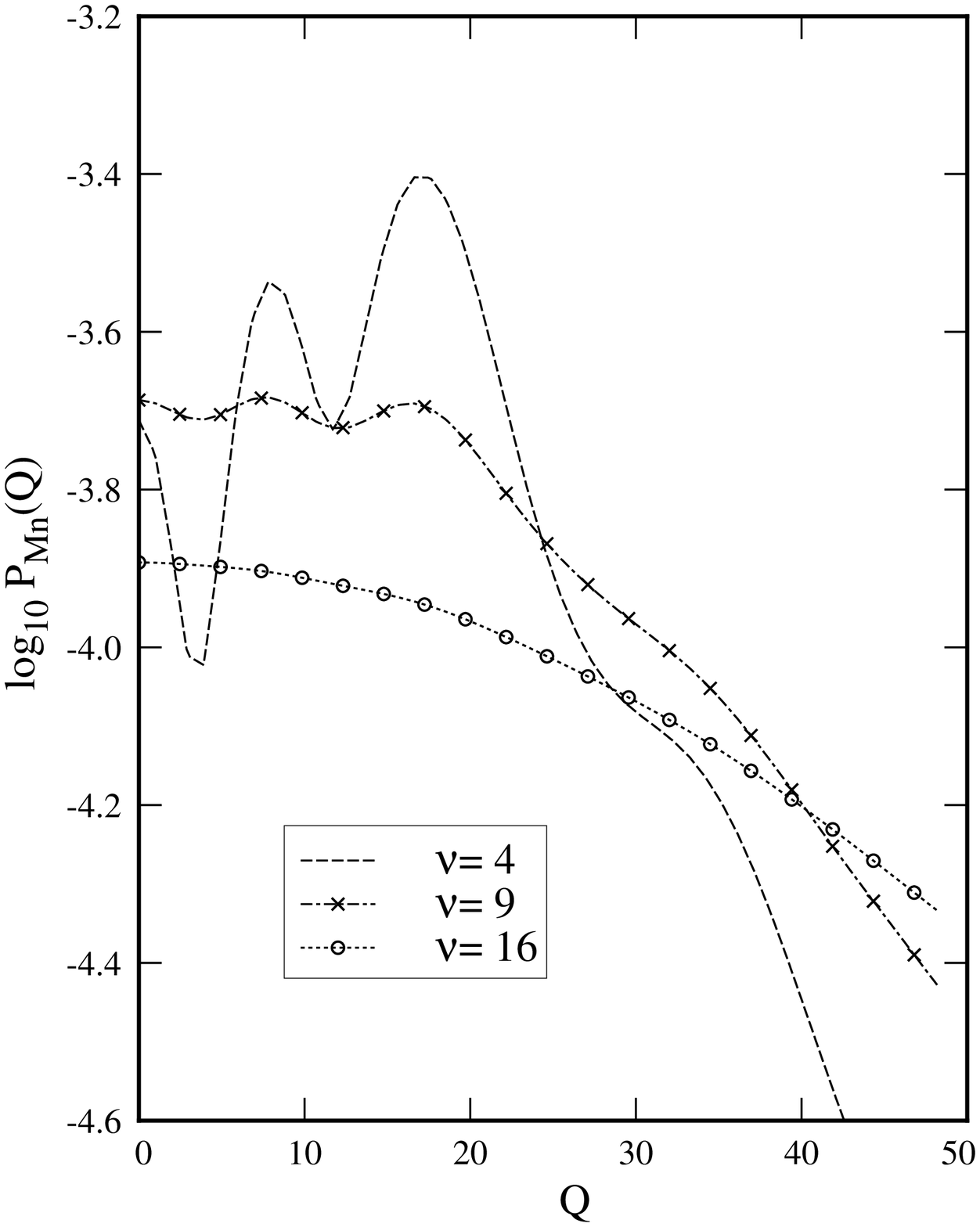} 
     }  
\caption{Dashed, linked-crosses  and linked-circles curves:   
$\log_{10}\, P_{Mn}(Q)$ for $\nu=4,$ $9$ and $16,$ respectively.}
\end{figure}
\begin{figure}[htb] \centering 
\mbox{  \epsfysize=120mm 
         \epsffile{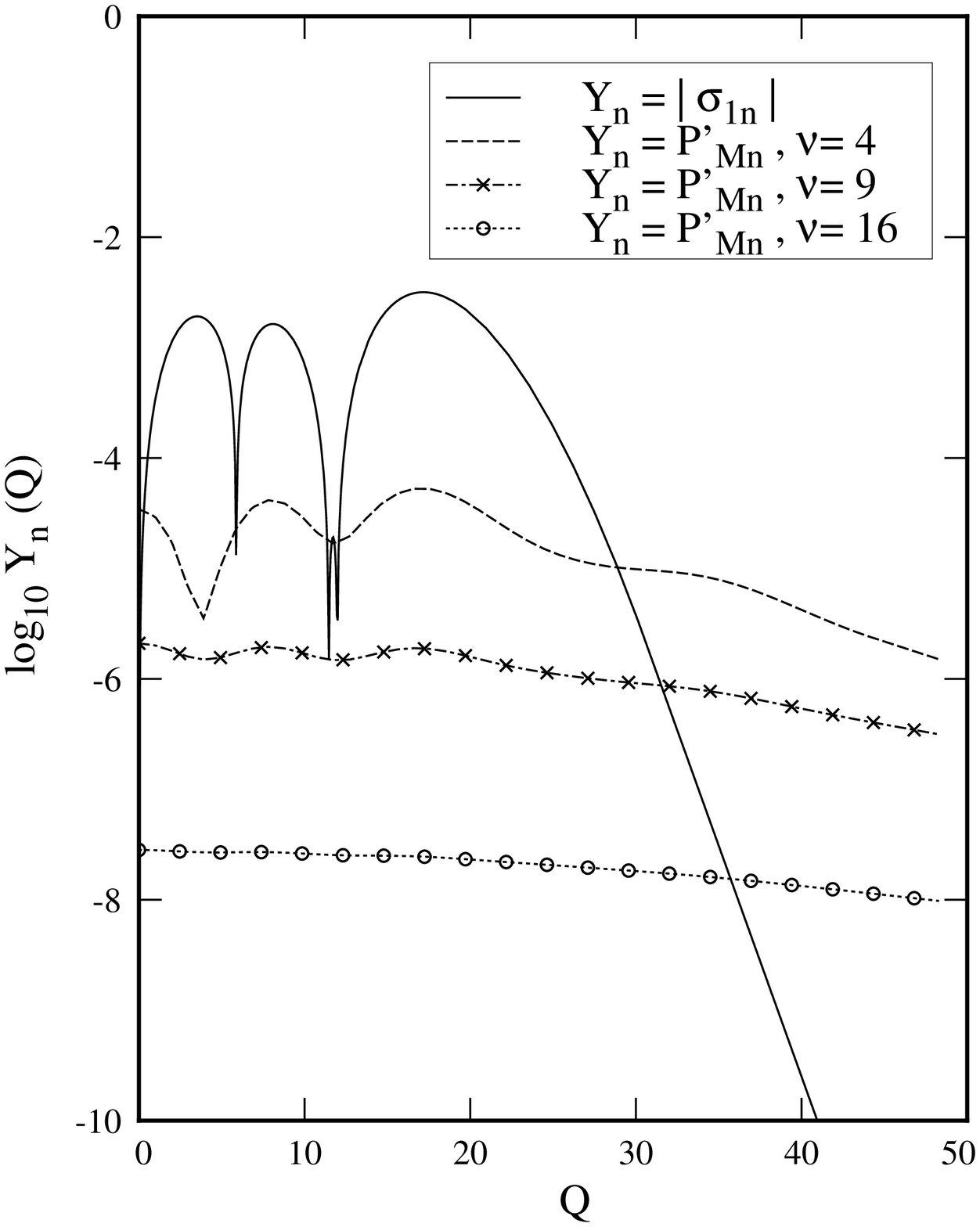} 
     }  
\caption{Solid curve: $\log_{10}\, |\sigma_{1n}(Q)|$ for a nuclear signal.
Dashed, linked-crosses and linked-circles curves: probability distributions 
$\log_{10}\, P^{\, \prime}_{Mn}(Q)$ when $\nu=4,$ $9$ and $16,$
respectively.}
\end{figure}
Besides the damping and information loss which are evident from Figs. 7 
and 8, we may stress a feature of Fig. 8, namely the transformation of 
``negative cross sections'' into positive ones after multiple scattering.
In order to create interferences, it was necessary, at the stage of
making  a model for $\sigma_{1n},$ to create negative values interfering
with $\sigma_{1c}.$ As has already been pointed out, this is allowed as
long as $\sigma_1$ remains positive; there is a degree of freedom in modeling
$\sigma_{1n}.$ The solid curve in Fig. 8 shows $\log_{10}\,  |\sigma_{1n}|.$
One sees four arches, the first and the tiny third ones meaning negative
values. Such ``negative'' arches disappear in the dashed curves representing
$P^{\, \prime}_{Mn}.$ This disappearance justifies the use of models where
$\sigma_{1n}$ can be not everywhere positive as long as $\sigma_{1n} + 
\sigma_{1c}$ is everywhere positive, as was discussed after
Eq. (\ref{mdlsig1n}).   

An advantage of our use of special analytical forms for the cross sections,
Eqs. (\ref{mdlsig1c},\ref{mdlsig1n}), is that such forms induce analytical
 profiles, Eqs. (\ref{mdlprf1c},\ref{mdlprf1n}), which in turn allow
 analytical forms for the multistep profiles, Eqs. (\ref{thrdstep}) and
(\ref{forward}-\ref{correctionC}). Values of $\langle Q^2 \rangle$ can then
be easily obtained from the use of Eq. (\ref{ruse}). The rates of broadening
as functions of $\nu$ can also be readily calculated.  Fig. 9 shows how, at
values of $\nu$  smaller by several orders of magnitudes than those estimated
from geometric cross sections, the square-widths $\langle Q^2 \rangle$ of
$P_{Mn}, P_{Mc}$ already increase linearly with $\nu.$ We have also noted a
similar behavior of the widths of $P^{\, \prime}_{Mn}$ and
$P^{\, \prime}_{Mc}$. 
\begin{figure}[htb] \centering 
\mbox{  \epsfysize=120mm 
         \epsffile{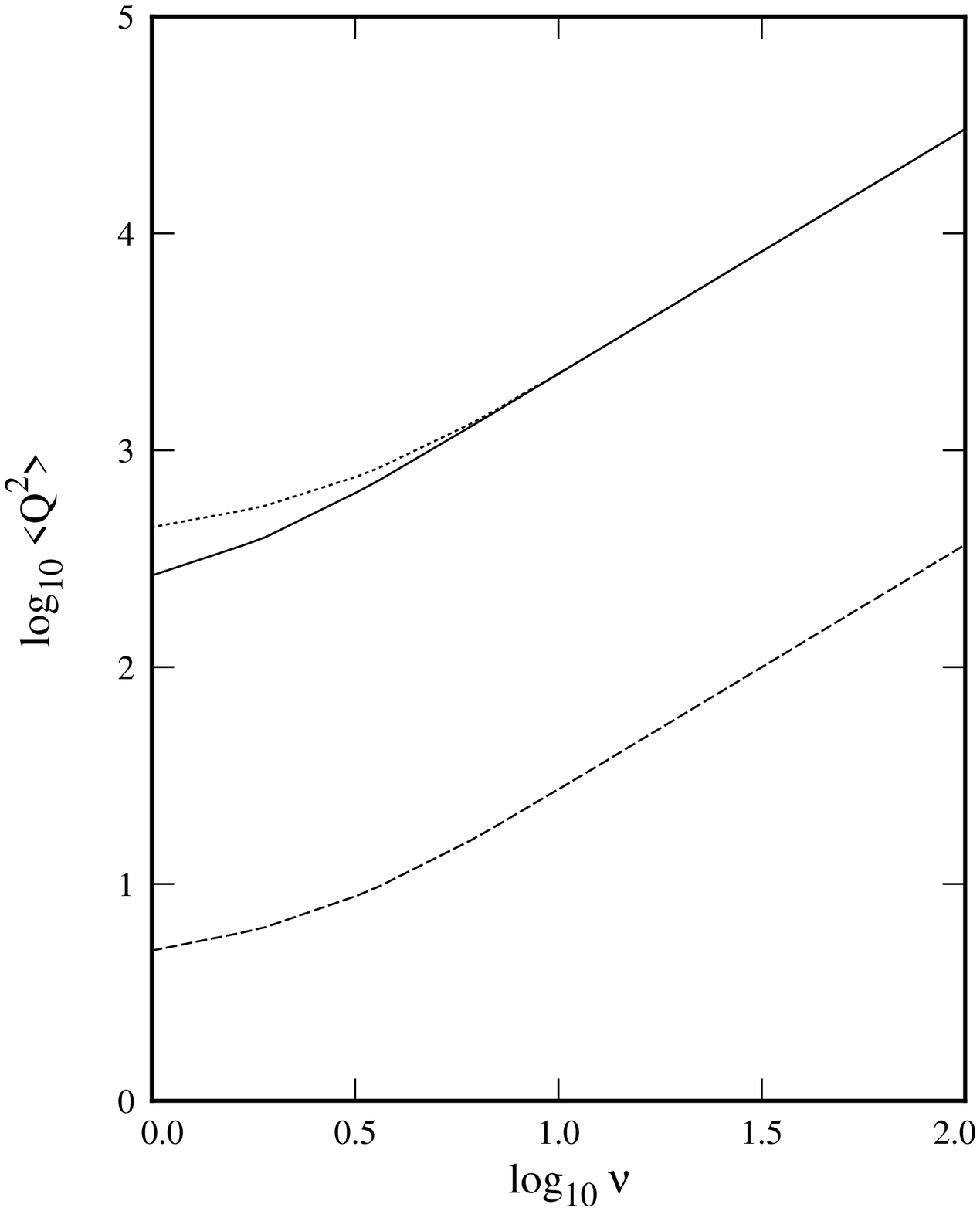} 
     }  
\caption{The dependences on $\log_{10}\, \nu$  of 
$\log_{10}\, \langle Q^2\rangle$   
for $P_M(Q)$ (solid curve), $P_{Mc}(Q)$ (dashed curve) and 
$P_{Mn}(Q)$ (dotted curve). Note that the slopes $\simeq 1$ when $\nu >> 10.$}
\end{figure}

In summary, the signature of nuclear information (diffractive oscillations
in the differential cross section) in the region of small momentum transfers  
is washed away by the broadening of the angular distribution. This happens
even with our model that has exaggerated nuclear cross sections. In the next
section, we examine if there exist momentum-transfer regions where the
multiple scattering of the proton does not completely blurr nuclear signals.

\section{Conditions for observing nuclear information}

Let $r \equiv q_{_{1/2}}/q_{_{min1}}$ be the ratio of the half-width of the
Coulomb peak to the momentum transfer at which the first minimum due to
nuclear diffraction is observed for a thin target. At high energies, both
$q_{_{1/2}}$ and $q_{_{min1}}$ occur at very small angles. Consequently,
$r = \theta_{_{1/2}}/\theta_{_{min1}}$ with the $\theta$'s being the
respective scattering angles corresponding to $q_{_{1/2}}$ and
$q_{_{min1}}$. From Eq. (\ref{broadenrule}), it is reasonable to expect 
a rule, $\langle q^2 \rangle_M \simeq \nu\, \langle q^2 \rangle_1
\simeq \nu\, q_{_{1/2}}^2$  $(= \nu r^2 q^2_{_{min1}}),$ hence that there is
a critical value $\nu_{crit} \simeq r^{-2},$ above which nuclear signals 
will be obliterated by the broadening of the Coulomb peak. In other words,
nuclear signals can only be observed at
$q \gg q_{min1}$ for $\nu > \nu_{crit}$. 

In Fig. 10, we show the elastic scattering differential cross sections of 
protons scattered from a thin $^{208}$Pb target at $\sim$20 GeV, 
which we have calculated by using the method of  
optical model of Ref.\cite{Matt} with      
a screened Coulomb interaction. The calculated cross sections
exhibit the main
characteristics of high-energy proton-nucleus scattering, namely, a narrow  
forward Coulomb peak and the diffractive oscillations at larger angles.   
Here, the diffractive pattern constitutes the nuclear signal.  One notes 
that the first diffractive
minimum lies at about $\sim 6$ milliradians. We have noted from our
calculation that $\theta_{1/2}$ for the Coulomb peak is of order
$\sim 0.003$ milliradians. It is therefore reasonable to assume that, for
high-energy proton scattering from nuclei, $r$ is of order $\sim 10^{-3}$ or
less in general. At most one might consider $r$ of order  $\sim 10^{-2}$.
Accordingly, although the geometric size of a nucleus is typically
$\sim 10^4-10^5$ smaller than that of its atom, the range of ``nuclear
information profiles'' at high scattering energies may be taken  
$\sim 10^2$ to $\sim 10^3$ smaller than the range of the screened atomic
profile, and possibly much smaller. While the model used in the previous
section where $\sim 0.5 <  r < \sim 0.1$ is pedagogically justified, an 
analysis with a smaller $r$ is in order.
\begin{figure}[htb] \centering 
\mbox{  \epsfysize=120mm 
         \epsffile{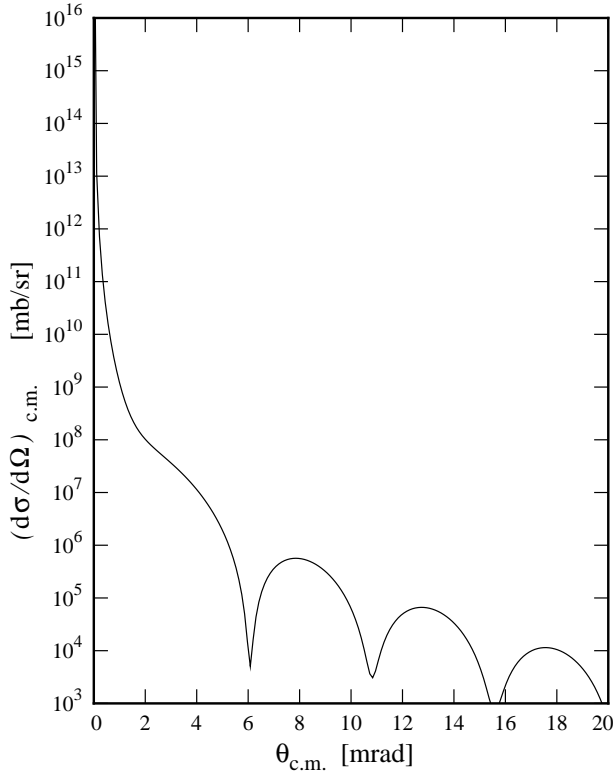} 
     }  
\caption{Differential cross sections 
of p$-^{208}$Pb elastic scattering at 20 GeV.}   
\end{figure}

We first introduce a model in which $r=10^{-2};$ the
profile function $\phi_1(B)$ is the sum of a ``Coulomb'' term,
\begin{equation}
\phi_{1c}(B)  = B\, K_1(B) 
\label{windowC} 
\end{equation} 
and a ``nuclear'' term,
\begin{equation}
\phi_{1n}(B)= \frac{4 \times 10^{-6}}{1 + \exp[800\, (B-1/100)]}\, .
\label{windowN}
\end{equation}  
The profile $\phi_{1c}$ gives a bare Coulomb cross section of the form
$\sigma_{1c}(Q) \propto (1+Q^2)^{-2}$ and the Woods-Saxon profile $\phi_{1n}$
makes, in practice, a window with range $r=1/100$ indeed. 
The coefficient $800$ in
its exponent creates a ``smoothed'' Heaviside function.  Both profiles are
normalized so that $\sigma_{1c}^{tot}=1,$ and
$\sigma_{1n}^{tot}/\sigma_{1c}^{tot}=4 \times 10^{-6}$ except for a
negligible factor $1+e^{-8}.$ This cross-section ratio is quite compatible
with the $r^2$ suggested by Fig. 10. Hence, the set of parameters given in
Eqs. (\ref{windowC}) and (\ref{windowN}) is more realistic than that used
in the previous section. The result for various angular cross sections
$\sigma_M(Q;\nu),$ compared with the single scattering $\sigma_1(Q),$ is
shown in Fig. 11. An inspection of the figure shows that the Coulomb peak
damps and spreads and the nuclear signal fades when $\nu$ increases. The solid  
curve, representing $\log_{10}\, \sigma_1(Q),$ and the dashed curve,
representing $\log_{10}\, \sigma_M(Q)$ for $\nu=2000,$ exhibit somewhat
similar oscillations. The dotted curve, corresponding to
$\log_{10}\, \sigma_M$ for $\nu=10^4,$ hardly oscillates any more, {\em i.e.,}
nuclear signals are completely washed out. This confirms a loss of nuclear
signal at low and moderate momentum transfers when $\nu$ approaches
$\nu_{crit} \sim r^{-2}= 10^4$.
\begin{figure}[htb] \centering 
\mbox{  \epsfysize=120mm 
         \epsffile{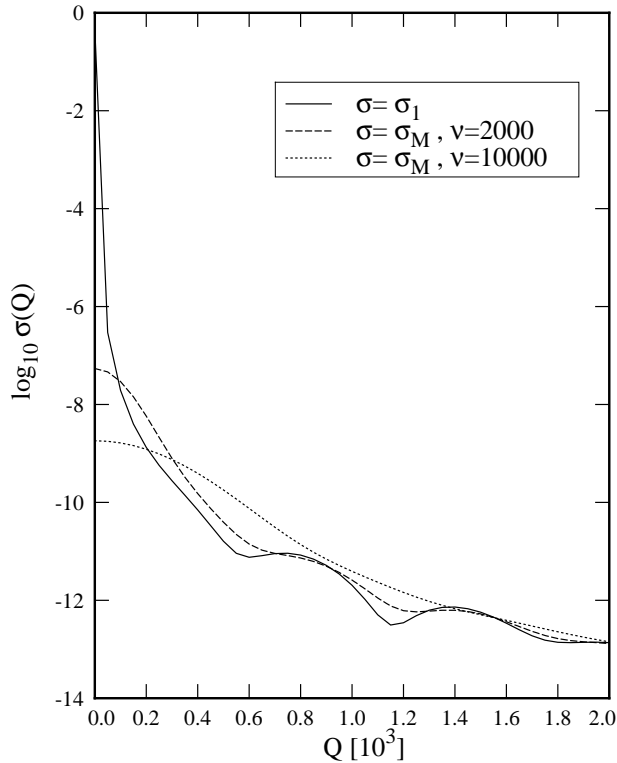} 
     }  
\caption{The dependences on $Q$ of $\log_{10}\, \sigma_1$ (solid curve) and 
of $\log_{10}\, \sigma_M$ for $\nu=2000$ (dashed curve) and 
$\nu=10000$ (dotted curve). All with $r=10^{-2}$.}
\end{figure}

The $\nu_{crit} \sim r^{-2}$ rule is also seen in the previous section, where 
the use of $\sim 0.5 < r < \sim 0.1$ induces the loss of nuclear signal as 
early as $\nu > \sim 10$. To further verify this rule, we use the same 
$\phi_{1c}$ but use instead $r= 0.5 \times 10^{-2}$ in $\phi_{1n}$, namely,  
\begin{equation}
\phi_{1n}(B)= \frac{4 \times 10^{-6}}{1 + \exp[1600\, (B-1/200)]}\ .  
\label{windowNN}
\end{equation} 
The results are shown in Fig. 12. As we can see, the observation of nuclear
signals is much improved; in agreement with the rule that $r=1/200$ elevates 
$\nu_{crit}$ to a higher value, $\sim 4 \times 10^4$. 
\begin{figure}[htb] \centering 
\mbox{  \epsfysize=120mm 
         \epsffile{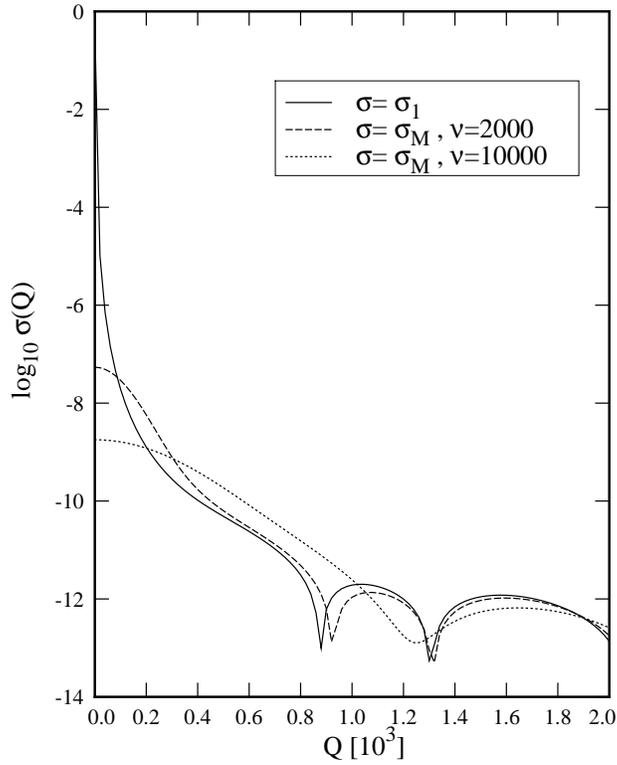} 
     }  
\caption{The dependences on $Q$ of $\log_{10}\, \sigma_1$ (solid curve) and 
of $Y=\log_{10}\, \sigma_M$ for $\nu=2000$ (dashed curve) and $\nu=10000$ 
(dotted curve). All with $r=0.5\times 10^{-2}$. }
\end{figure}
 
As a last test of the $\nu_{crit}\sim r^{-2}$ rule, 
we keep $\phi_{1c}(B)=B K_1(B)$ and let  
\begin{equation}
\phi_{1n}(B)= \frac{2 \times 10^{-8}}{1 + \exp[8000\, (B-1/1000)]}\ ,   
\end{equation}
which is an obvious $r=1/1000$ case. The results are shown in 
Fig. 13, where it is clear that, as expected,
$\nu_{crit}$ occurs between $10^{-5}$ and $10^{-6}.$

\begin{figure}[htb] \centering 
\mbox{  \epsfysize=120mm 
         \epsffile{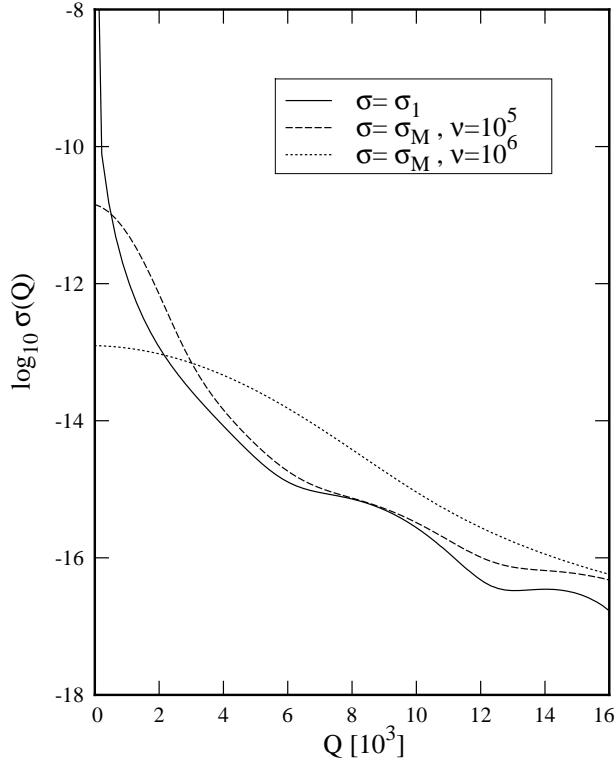} 
     }  
\caption{The dependences on $Q$ of $\log_{10}\, \sigma_1$ (solid curve) 
and of $\log_{10}\, \sigma_M$ for $\nu=10^5$ (dashed curve) 
and $\nu=10^6$ (dotted curve). 
All with $r=10^{-3}$. }
\end{figure}
Because $\nu = T \rho\, \sigma_1^{tot} \simeq T \rho\, \sigma_{1c}^{tot}$,
the existence of a $\nu_{crit}$ induces a critical target thickness $T_{crit}$
such that the retrieval of nuclear signal is possible for target thickness
$T$ sufficiently less than $T_{crit}$; namely, 
\begin{equation} 
T < T_{crit} \simeq \frac{\nu_{crit}}{\rho\, \sigma_{1c}^{tot}}\ . 
\label{Tmax} 
\end{equation} 
For p$-^{208}$Pb elastic scattering at 20 GeV, $\sigma_{1c}^{tot}\simeq 6.7 
\times 10^8$ mb $=6.7 \times 10^{-19}$ cm$^2$. The density $d$ and the atomic
mass number $A$ of lead are 11.3 g/cm$^3$ and 208, respectively. Hence,
$\rho =(d/A) {\mathcal N}_{Avog} = 3.27 \times 10^{22}$ cm$^{-3}$, with 
the Avogadro number ${\mathcal N}_{Avog} = 6.02 \times 10^{23}$ $[$1/mole$]$. 
Because from Fig. 10 it is likely that $r\sim 10^{-3}$ and our analysis
indicates that $\nu_{crit} \sim 10^6$, then Eq. (\ref{Tmax}) gives
$T_{crit}\simeq 46$\ cm. Hence, a nuclear signal can be retrieved at
$\nu\leq 10^5,$ which corresponds to $T\leq 4.6= 0.1\, T_{crit}\ [$cm$]$. In
so far as $Q=1$ corresponds to $\theta \sim 0.003$ milliradians, the 
survivor oscillation seen for $\nu=10^5$ in Fig. 13 between $Q \sim 6000$ and
$Q \sim 12000,$ compatible with the expected period $\sim 2 \pi/r,$ would 
demand experimental measurements at angles of order a few dozens of 
milliradians at most.

The proton-nucleus Coulomb cross section $\sigma_{1c}^{tot}$ is 
$\propto Z^2 R^2_e \sim Z^{4/3}$, where $Z$ is the target charge and 
$R_e$ is the root-mean-square radius of electric charge distribution in an 
atom with $R_e \sim a_0/Z^{1/3}$ and $a_0$ being the first 
Bohr radius\cite{Gott}. Hence, one can estimate $\sigma^{tot}_{1c;pA}$ for
proton scattering from a given nucleus $A$ at 20 GeV by using    
$\sigma^{tot}_{1c;pA} \sim \lambda\, \sigma^{tot}_{1c;p-Pb} $ with the
scaling factor $\lambda = (82/Z)^{-4/3}$. Hence, for a same $\nu$ one has 
$T(pA) = T(pPb)\rho_{Pb}/(\rho_{A}\lambda)$. Results for a sample of atomic
nuclei at $\nu = 10^5$ are given in Table I.
Of course, the price one pays in studying the nuclear signals that
survive the multiple-scattering broadening is that one has to measure the
angular distribution with good energy resolution
at large proton scattering angles. In the case of 
20 GeV incoming protons, the angles are about tens of
milliradians, where the magnitudes of the cross sections are quite
small. However, such measurements should be feasible with the currently
available technology. 
\begin{table}
\begin{center}
\caption{ Target thickness $T$ corresponding to $\nu=10^5$ for 20-GeV protons.}
\label{tabT}
\begin{tabular}{|c|c|c|c|c|c|c|} \hline 
Atom & d\ \ [g/cm$^3$] & Z & $\lambda$ & A & $\rho$\ \ [10$^{22}$\ cm$^{-3}$]
& $T$\ \ [cm] \\ \hline \hline 
Pb & 11.3 & 82 & 1.000 & 208 & 3.27  & 4.6  \\ \hline
Cu &  8.9 & 29 & 0.250 &  64 & 8.39  & 7  \\ \hline
Al &  2.7 & 13 & 0.086 &  27 & 6.02  & 29 \\ \hline 
Mg &  1.74 & 12 & 0.077 & 24 & 4.37  & 44  \\ \hline
Be &  1.85 & 4 & 0.018  &  9 & 12.38 & 67  \\ \hline  
\end{tabular}
\end{center}
\end{table}
\nopagebreak

\section{Conclusion}

The main mathematical and physical statement proposed by this work about
multiple scattering consists in folding probabilities rather than scattering
amplitudes. This is justified by the incoherence which is expected between the
different scatterers of a thick, non crystalline target. Simultaneously, an
eikonal approximation, justified by the very high energy of the beam, allows
a familiar impact parameter representation with profiles. Furthermore,  
small-momentum expansions\cite{Moli} are not employed in the formulation.  
As a consequence of such initial statements, a Poisson process is found,
leading to an elementary formalism of convolutions and exponentiations in a
context of Fourier-Bessel transforms.

This Poisson process is nothing but a random walk in transfer momentum space.
The central limit theorem is at work and the details of nuclear oscillations
and interferences between Coulomb and nuclear scattering are blurred very
fast as soon as the parameter $\nu=T \rho\, \sigma_1^{tot},$ a measure of the
number of collisions, exceeds a critical value of order $r^{-2}.$ Here $r$ is 
the ratio of the range of the nuclear profile to that of the atomic profile.

Below this critical value of $\nu,$ and at moderate and large momentum
transfers (at the cost of very small elastic cross sections in the latter case)  
our conclusion is that some nuclear information remains observable. Such 
information is contained in oscillations of the multistep angular cross
section $\sigma_M(Q)$ with periods $\sim 2 \pi/r,$ oscillations that are
similar to those of the Bessel function, $J_1(r\, Q),$ which typically
represents pure nuclear diffractive scattering. Our model analysis shows that
the characteristic distances between successive cross-section maxima and
minima in the angular distribution remain essentially unchanged while each of
these oscillations dampens as $\nu$ increases. 

From the point of view of retrieving nuclear signals from protons traveling 
through a thick target, for which no sufficient attention was given in the 
literature, our work is more of a general feasibility study 
rather than a specific numerical evaluation. We have made use of analytical
and semi-analytical models to bring out the basic features of the underlying
physics. We believe that the positive feasibility concluded from this study
will sustain tests in detailed numerical applications.  
  
One problem which has not been solved in the present work, however, is to find
an estimate of the $\nu$ dependence of such periods $\sim 2 \pi/r.$ Our 
numerical evidence suggests that the dependence is not very strong, despite all
the causes for a broadening of the signal, but our models and calculations lack
the precision needed to tabulate such periods into functions of $\nu.$ This
effort is under consideration for an extension of the present work.
 
In summary, below $\nu_{crit}$, which is of order $1/r^2$ 
with $r$ being the ratio of the range of nuclear profile to the range of 
atomic profile, the nuclear signals can be retrieved from proton scattering  
from a thick target of thickness $T < T_{crit}$. We suggest   
a conservative upper bound, namely, $T \leq 0.1\ T_{crit}$ for practical 
considerations.  
  
\vspace{1cm} 


\end{document}